\documentclass[amsfonts,twocolumn,showpacs]{revtex4}
\usepackage{epsfig,amsmath,amssymb,bm,graphics,psfrag,verbatim,subfigure}
\usepackage[usenames,dvipsnames]{color}

\newcommand{\be}{\begin{eqnarray}}
\newcommand{\ee}{\end{eqnarray}}
\newcommand{\bw}{\begin{widetext}}
\newcommand{\ew}{\end{widetext}}
\newcommand{\Tr}{{\rm{Tr}\,}}
\newcommand{\nh}{{\hat{n}}}

\begin{document}

\title{Morphology of Nematic and Smectic Vesicles}

\author{Xiangjun Xing$^{1}$}
\author{Homin Shin$^{2}$}
\author{Mark J. Bowick$^{3}$}
\author{Zhenwei Yao$^{3}$}
\author{Lin Jia$^{4}$}
\author{Min-Hui Li$^{4}$}
\affiliation{%
$^1$Institute of Natural Sciences and Department of Physics,
Shanghai Jiao Tong University, Shanghai 200240 China\\
$^2$Department of Polymer Science and Engineering, University of Massachusetts, Amherst, MA 01003 \\
$^3$Physics Department, Syracuse University, Syracuse NY 13244\\
$^4$Institut Curie, CNRS, Universit\'{e} Pierre et Marie Curie, UMR168, Laboratoire Physico-Chimie Curie, 26 rue d'Ulm, 75248 Paris CEDEX 05, France
}

\date{\today}


\maketitle

{\bf
 Recent experiments on vesicles formed from block copolymers with liquid-crystalline side-chains reveal a rich variety of vesicle morphologies.  The additional internal order (``structure") developed by these self-assembled block copolymer vesicles can lead to significantly deformed vesicles as a result of the delicate interplay between two-dimensional ordering and vesicle shape.  The inevitable topological defects in structured vesicles of spherical topology also play an essential role in controlling the final vesicle morphology.  Here we develop a minimal theoretical model for the morphology of the membrane structure with internal nematic/smectic order. Using both analytic and numerical approaches, we show that the possible low free energy morphologies include nano-size cylindrical micelles (nano-fibers), faceted tetrahedral vesicles, and ellipsoidal vesicles, as well as cylindrical vesicles. The tetrahedral vesicle is a particularly fascinating example of a faceted liquid-crystalline membrane.  
Faceted liquid vesicles may lead to the design of supra-molecular structures with tetrahedral symmetry and new classes of nano-carriers.
}

Amphiphilic block copolymers in water, like natural phospholipids, can self-assemble into various monolayer or bilayer structures, such as micelles and vesicles~\cite{discher1999,discher2002}.  In particular, rod-coil block copolymers, with a flexible hydrophilic chain and one or more rod-like hydrophobic blocks, exhibit a rich morphology of structures, and therefore have significant potential to advance fundamental science and drive technological innovations~\cite{halperin,jenekhe, jenekhe2,nowak,bellomo,palmer,wang,he,yang,mabrouk}.
Among these rod-coil block copolymers, we are especially interested in liquid crystalline (LC) block copolymers in which the hydrophobic block is a nematic or smectic liquid crystal polymer~\cite{pinol,jia,xu,boisse,barrio,yangh,jia2,jia3}.  The self-assembly process of LC block copolymer vesicles is not completely controlled by the energetic and entropic interactions between different parts of the polymers.  The in-plane LC order and the associated defect structure also play very important roles in determining the intermediate and final shape of vesicles.  The tailor-design of both material properties and vesicle morphology by controlling the molecular structures of the block polymers is state-of-the-art research in the fields of polymer science, materials science and chemical engineering.

Some of the structures formed by these LC side-chain block copolymers in aqueous solution are rather counterintuitive, such as faceted vesicles, nanotubes and compact vesicles with tiny inner space~\cite{xu, jia3}.  In all these structures, the in-plane smectic order is clearly visible under Cryo-TEM.   In this article we develop a theoretical explanation of the geometric structures of vesicles with in-plane nematic or smectic order.  We present a simple model free energy as a functional of both the membrane geometry and the in-plane nematic order.  Using both analytic and numerical methods, we then analyze the low free energy morphologies in various parameter regimes.

Focusing on their overall shape we first look at the model free energy of a self-assembled monolayer as a functional of their shape and nematic order parameters~\cite{mackintosh,frank}:
\be
H_{\rm m} &=& \frac{1}{2}  \int \sqrt{g} \, d^2 x
\left[  K \, (\vec{D} \nh)^2
+  \kappa \, (H - H_0)^2
\right] 
 \label{H-monolayer}
\ee
Here $K$ is the Frank constant in the one-constant approximation, while $\vec{D}$ denotes the covariant derivative.  $H$ is the mean curvature and $H_0$ is the spontaneous curvature, which is determined by the asymmetry in the sizes of the hydrophobic and the hydrophilic parts of the LC block copolymers.  We shall choose the normal vector of the monolayer to point from the hydrophobic side to the hydrophilic side.  Therefore $H > 0$ means that the hydrophilic side is bending outwards.


All three parameters $K, \kappa, H_0$ depend on the chemical structures of the block copolymers as well as their interaction with the solvent in a complicated way.  Furthermore, strictly speaking, a nematic membrane is locally anisotropic.  Therefore its Frank free energy is characterized by two constants: one for splay ($K_1$) and one for bend ($K_3$).  Likewise, the bending energy as well as the spontaneous curvature should also be generically anisotropic, characterized by three bending constants and three spontaneous curvature components.  Such a model is characterized by 8 independent parameters and is extremely complicated to analyze.  For the sake of simplicity, we shall focus on the greatly simplified toy model Eq.~(\ref{H-monolayer}), which captures the essential physics of nematic vesicles which is the competition between the extrinsic bending energy and the two-dimensional Frank free energy.

A more important, conceptual issue is the following: In what sense can the vesicle morphology be understood in terms of minimization of elastic free energy Eq.~(\ref{H-monolayer})?  As is well known, the formation of vesicles is a complicated nonequilibrium process.  Whether a certain property of a vesicle is distributed according to Gibbs-Boltzmann depends on the relevant experimental time scale, and on the time scale at which the given property equilibrates.  At the stage of vesicle formation, individual molecules on the membrane can diffuse quite efficiently.  Motion of liquid crystalline defects, however, requires coherent movement of all polymers on the vesicle, and is usually very slow.  Hence we expect that the vesicle morphology achieves a local thermal equilibrium, where the shape and LC order minimizes the elastic free energy Eq.~(\ref{H-monolayer}) (with appropriate parameters corresponding to the physical conditions under which the self-assembly takes place),  subject to global constraints of given vesicle topology and LC defects distribution.   We shall then enumerate all possible vesicle topologies and compare these free energy minima.   It is interesting to note that in recent experiments by Jia {\it et. al.}~\cite{pinol}, multiple vesicle topologies were often observed using a given preparation method, suggesting that kinetics of self-assembly also played an important role in the selection of vesicle morphology.  

Smectic vesicles can also be viewed as nematic vesicles with bending constant much larger than splay constant.  On a membrane with in-plane smectic order, therefore, the bending deformation of the nematic director field should vanish everywhere.  Mathematically this is equivalent to $\nh \cdot D \nh = 0$, that is, the nematic director locally follows the geodesics.  This is always possible, for an arbitrary but prescribed membrane shape, except at the core of nematic disclinations.  For these configurations, the Frank free energy becomes independent of the bending constant. Hence Eq.~(\ref{H-monolayer}) is also a toy model for membranes with in-plane smectic order, with the understanding that $K$ is the splay constant and the nematic director strictly follows the local geodesics.

\begin{figure}
	\centering
		\includegraphics[width=8cm]{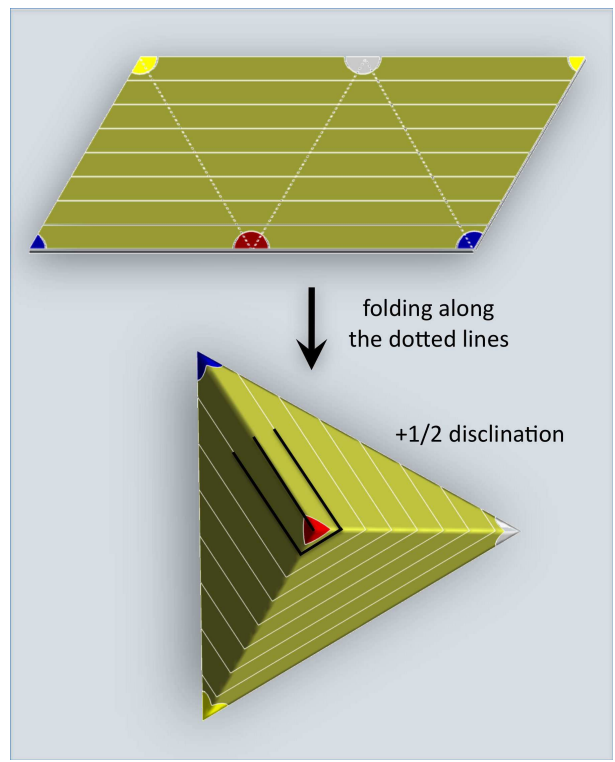}
	\caption{Top: A tetrahedron compatible with $+1/2$ disclination on each vertex can be constructed using a parallelogram, by folding along the dashed lines. A constant nematic director field in the unfold parallelogram is shown by the array of parallel straight lines. After folding up, the angles with same color circle around the same vertex. Bottom: the tetrahedron obtained via folding. There is exactly one $+1/2$ disclination on each vertex of the tetrahedron. }
  \label{fig:tetrahedron-schematics}
\end{figure}

Minimization of the Frank free energy in Eq.~(\ref{H-monolayer}) requires that the covariant derivatives of the nematic director field vanish everywhere on the surface.   As is well known in differential geometry, this is possible only if the Gaussian curvature vanishes everywhere, i.e. the surface is a {\em developable surface}.  The family of developable surfaces includes planes, cylinders, cones, and tangent developable surfaces \footnote{Surfaces spanned by tangent lines of a spatial curve: a generic tangent developable surface that is topologically identical to plane is however not expect to be observed in experiments, because it can easily relax to a plane which has bending energy.  It is not clear to us whether there exists a tangent developable surface that is topologically different from all the aforementioned structures.}.  On the other hand, minimization of the bending energy in Eq.~(\ref{H-monolayer}) leads to a constant mean curvature $H_0$.  It is clear that the only geometry minimizing both terms in the free energy in Eq.~(\ref{H-monolayer}) is a cylindrical monolayer with a given radius $1/H_0$.  In the recent example of Jia {\it et. al.}~\cite{pinol}, for example, where only aqueous solvent is present at the final stage of assembly, monolayer cylinders with very small radius (nanofibers) are observed.  
The inner space of the cylinders is completely filled by the hydrophobic parts of the polymers.  In order to form monolayer cylinders with larger radius, the inner space has to be filled by solvent (or other polymers) that are friendly to LC blocks.  If there is only aqueous solvent, and if $1/H_0$ is not small, monolayer cylindrical structures with favorable spontaneous curvatures cannot pack space and therefore the system should form certain kinds of bilayer structures, where two monolayers with opposite orientation stack together.

The free energy of a bilayer membrane can be obtained by adding up the free energies for two monolayers on both sides of the bilayer:
\be
H_{\rm m} &=& \int \sqrt{g} \, d^2 x
\left[  K \, (\vec{D} \nh)^2
+  \kappa \, H^2
\right]
 \label{H-bilayer}
\ee
We shall focus on the morphology of bilayers in the remainder of this article.

Without considering the boundary effects, a flat bilayer with uniform nematic order clearly minimizes both terms in Eq.~(\ref{H-bilayer}). The energy cost associated with the boundary, however, increases with the system size, and  exceeds that associated with a closed vesicle with nonzero curvature, for sufficiently large systems \cite{helfrich}.  Close vesicles therefore must form for sufficiently large bilayer membranes.

The morphology of a bilayer is controlled by the competition between the extrinsic bending energy and the Frank free energy.  We shall first limit the discussion to closed vesicles of spherical topology.   Since the total Gaussian curvature is nonvanishing, the system is frustrated and the Frank free energy competes with the bending energy.  First consider the limiting case $K \ll \kappa $.  The dominant contribution to the total energy is then the bending energy: minimizing this leads to a round spherical shape.  For a more realistic model where the bending energy is not isotropic, however, the shape will reflect the anisotropy of the bending moduli, leading to ellipsoidal shapes.  The exact form of the shape as a function of the bending moduli is rather difficult to calculate, however, and will not be treated in this article.  Ellipsoidal vesicles are frequently observed in the experiments of Jia {\it et. al.}~\cite{jia}, with the smectic layers all perpendicular to the long axis of the ellipsoid.  It can be inferred from this observation that the bending rigidity is higher along the nematic director than perpendicular to the director.

Let us now consider the opposite regime where $K \gg \kappa$.  In this case, the system should first minimize the Frank free energy, which leads to developable surfaces with vanishing Gaussian curvature everywhere.  This is clearly not possible due to Gauss' {\em Theorem Egregium}, which states that the total integrated Gaussian curvature of a surface with spherical topology is a topological invariant and equal to $4 \pi$.  There are faceted polyhedral surfaces, however, for which the Gaussian curvature vanishes everywhere but at a discrete number of (singular) vertices.  These vertices are the ideal locations for orientational defects of the LC order (misery loves company)~\cite{LP:1992}. The total defect strength on a closed surface is also a topological invariant, according to the Gauss-Bonnet theorem.  For nematic and smectic orders, this theorem dictates that on a sphere (or any other surface with the same topology), there are three possibilities for the structure of defects: 1) four disclinations each with strength $+1/2$; 2) two defects each with strength $+1$; 3) one strength $+1$ defect and two strength $+1/2$ defects.  Now one needs at least four points to span a non-degenerate polyhedron; a tetrahedron in the minimal case.  We conclude that in the limiting case $K \gg \kappa$, the ground state morphology of a vesicle with spherical topology is a faceted tetrahedron, with a strength $1/2$ disclination located at each of the four corners.  This structure is indeed observed in recent experiments \cite{xu, jia3}, as well as in our simulation, to be discussed in detail below.

It is important to note that not all tetrahedra support a suitable nematic defect configuration. To ensure that the director field has vanishing covariant derivative everywhere except at the four vertices, but including the six edges, the sum of the three angles surrounding every vertex of the tetrahedron must be $180^{\circ}$.  This imposes three constraints on the geometry of the tetrahedron. \footnote{Naively we see there are four constraints but the condition for one vertex follows automatically from the constraints for the other three.}  Since the set of all tetrahedral shapes (up to scaling the overall size) forms a five dimensional space, we see that the set of all fixed-size tetrahedra with vanishing covariant derivative everywhere except the vertices forms a {\em two dimensional manifold}.   Fig.~\ref{fig:tetrahedron-schematics} illustrates how these tetrahedra, together with a nematic director field with vanishing covariant derivative, can be constructed by folding a parallelogram.  These tetrahedra have the special property that all four triangular facets are identical.  All these structured smectic vesicles have vanishing Frank free energy.  The degeneracy is lifted by different bending energies.  The total free energy of the system is given by the sum of the bending energies localized on the six edges and the defect core energies localized at the four vertices.  It is rather easy to see that for a given total surface area, the regular tetrahedron has a minimal value for the sum of all edge lengths.  Thus {\em the ground state morphology of  a smectic vesicle with spherical topology is a regular tetrahedron when the bending rigidity is vanishingly small}.   The transition between different shapes is probably extremely slow, however, as it requires coherent motion of all four nematic disclinations together with the overall smectic layer texture.

The edges and corners cannot be infinitely sharp in a realistic system.  They are rounded by either the membrane thickness, the core size of a nematic defect or the small bending rigidity $\kappa$.  Likewise, the bending energy on the edges must be finite. In a realistic self-assembly process, the bending energy may also be partially relieved by preferential aggregation of large polymers on the outside and smaller polymers on the inside of the membrane near the ridges and corners.  Faceted surface structures were studied previously in large viral capsids~\cite{lidmar,bruinsma}, which are formed by crystalline packing of proteins.  There the faceting is energetically favorable because it reduces the in-plane strain energy of the crystalline order formed by the constituent proteins.  What we have shown here is that a similar faceting can also be driven by the Frank free energy of LC order, despite their liquid nature.

\begin{figure*}
\centering
\includegraphics[angle=0.0, scale=0.115]{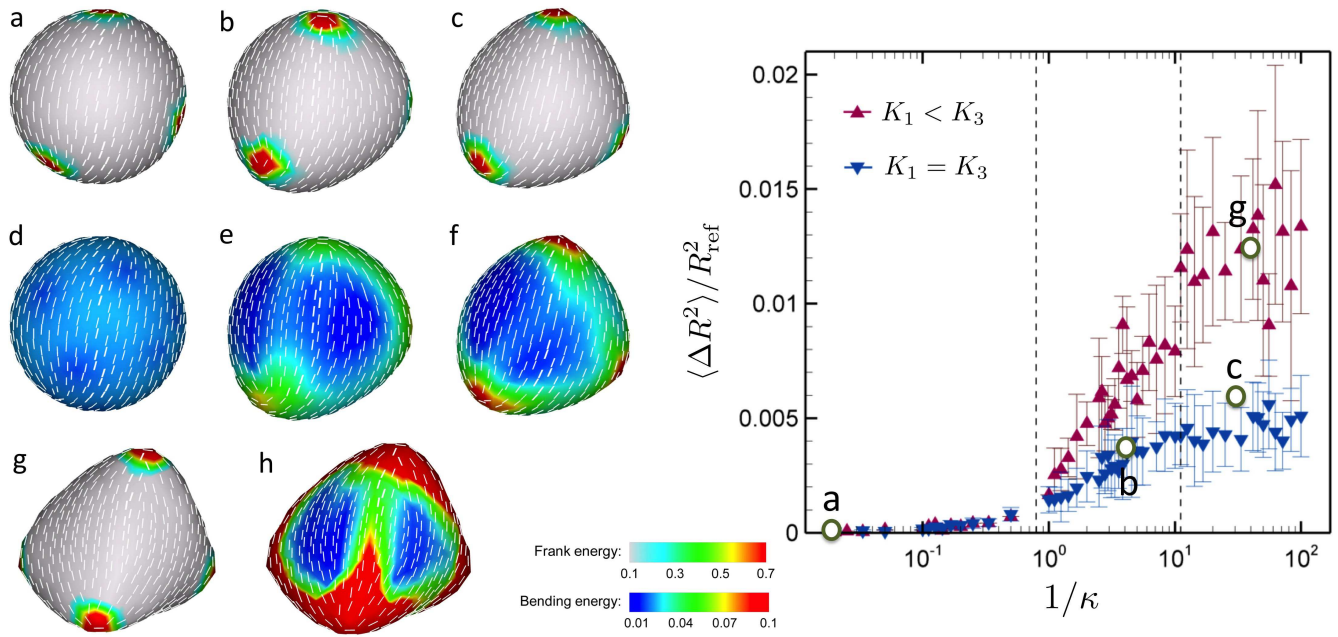}
\caption{ {\bf Morphology of nematic vesicles at different bending rigidities.} Left: The local Frank ({\bf a}--{\bf c}, {\bf g}) and bending ({\bf d}--{\bf f}, {\bf h}) energy contour plots; for a case of $K_1 = K_3=1$, {\bf a} \& {\bf d}, $\kappa = 100.0$; {\bf b} \& {\bf e}, $\kappa = 0.3$; {\bf c} \& {\bf f}, $\kappa= 0.05$; for a case of $K_1 \not= K_3$, {\bf g} \& {\bf h}, $\kappa = 0.04$. The total Frank free energies (in units of $k_BT$ )are 62.83, 59.06, 57.1, and 65.27, respectively. The normalized total bending energies (in units of $k_BT$) are 12.55, 16.0, 18.58, and 50.34, respectively.  Right: Calculated asphericities of nematic vesicles as a function of the inverse bending rigidity of $1/\kappa$.  The inverse triangles are for a case of $K_1 = K_3$ and the triangles are for a case of $K_1 \not= K_3 (K_3/K_1 \approx 2.0)$.  The empty circles represent locations corresponding to the morphologies of {\bf a} (\& {\bf d}), {\bf b} (\&  {\bf e}), {\bf c} (\& {\bf f}), and {\bf g} (\& {\bf h}). }
\label{simulation}
\end{figure*}

Another candidate for a low free energy morphology is a long cylinder of double layers (nanotube), for which the Frank free energy also vanishes.  The bending energy is approximately given by
\be
H_{\rm nanotube} = \kappa {A}/{a^2}, 
\ee
where $a$ is the radius of the cylinder.  The total bending free energy is therefore linear in the membrane area.
The faceted tetrahedron, on the other hand, has the total free energy
\be
H_{\rm tetrahedron} = 4 \kappa L/b,
\ee
where $L$ is the length of ridges and $b$ is the radius of curvature of rounded out ridges.  Since the area of a tetrahedron grows quadratically in L, it follows that the total bending energy for a tetrahedron scales as the square root of the membrane area.   Large faceted tetrahedral vesicles thus have lower free energy than nanotubes.  Both morphologies, however, have been observed experimentally~\cite{jia,xu,jia3}.  Selection of vesicle morphology is also affected by kinetics of self-assembly, as we discussed above.

In order to quantitatively investigate the ground state morphology of nematic vesicles, we develop a lattice nematic model on a deformable surface with spherical topology and perform energy minimization by the method of simulated annealing Monte Carlo (MC).  Details of the discretized form of the free energy, whose continuum limit is given by eq.~(\ref{H-bilayer}), can be found in ref.~\cite{shin} and the Methods section.

The simulation results for nematic vesicles at various bending rigidities show remarkable morphological transitions, as displayed in Fig.~\ref{simulation}.  As the bending rigidity $\kappa$  decreases, the vesicles with an isotropic Frank elastic constant undergo substantial shape deformation: 1) the spherical morphology is found to be stable at large $\kappa$ (Fig.~\ref{simulation}a, d); 2) ridges connecting four defects develop as $\kappa$ becomes smaller than 1 (Fig.~\ref{simulation}b, e); 3) a tetrahedral vesicle forms at a vanishingly small $\kappa=0.05$  (Fig.~\ref{simulation}c, f).   The faceting transition occurs near $\kappa \simeq 1$.  The stable morphologies are determined by a delicate balance between the in-plane Frank energy and the bending energy as the surface deforms away from round.  Indeed, as $\kappa$ decreases from 100.0 to 0.05,  the Frank free energy falls from 62.83 to 57.1 at the expense of bending energy which increases from 12.55 to 18.58. The Frank energy is localized near the four defects, which consequently induce deformation around the vertices.  We note that while we continuously vary the bending rigidity $\kappa$ ranging from 0.01 to $100.0k_B$T, the isotropic Frank elastic constant $K$ is set to 1$k_B$T.  Therefore, our simulations are entirely consistent with our prediction that spherical vesicles are stable in the regime of $K \ll \kappa $, whereas the faceted tetrahedral vesicles become stable in the other extreme $K \gg \kappa$.

We also explore the effect of anisotropy in the Frank elastic constants by studying the regime in which splay dominates over bend.  The smectic regime, as noted earlier, corresponds to the limit $K_3 \gg K_1$.  In the current simulation, the anisotropy is estimated to be $K_3/K_1 \simeq 2.0$ (see supplementary information).  Although the shape transition trends are qualitatively similar for both the one-Frank constant and anisotropic cases, the anisotropy leads to a more dramatic shape transition, resulting in a considerably more faceted tetrahedral vesicle at a very small $\kappa=0.04$, as displayed in Figs.~\ref{simulation}g and h.   In fact, the splay dominant nematic texture enhances the faceting more than the isotropic case does.  This is clearly understood by considering two membranes which possess a +1 disclination defect with pure splay and pure bending nematic textures, respectively.  The pure splay always decreases the Frank energy by buckling out-of plane, because it allows the defect to escape into the third dimension and thus better align the nematic directors.  On the other hand, such out-of-plane deformation of the pure bending does not alter the Frank energy and therefore, the faceting of pure bending membranes is not favorable upon deformation.  Note that we are restricting ourselves here to the case of isotropic bending rigidity. 

\begin{figure}
\centering
\includegraphics[angle=0.0, width=7cm]{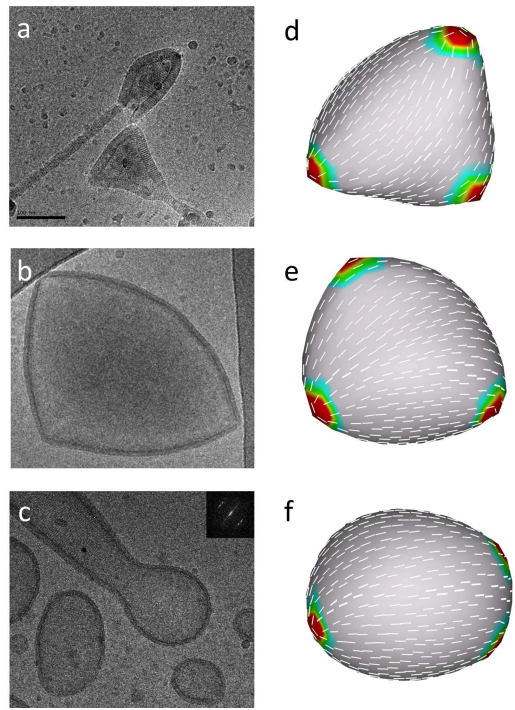}
\caption{ {\bf Comparison between experimental observations ({\bf a}--{\bf c}) and computer simulations ({\bf d}--{\bf f}).} Left: Experimental results. {\bf a}, a tetrahedron-shape smectic vesicle~\cite{jia3}; {\bf b}, a fat tetrahedron-shape smectic vesicle~\cite{xu}; {\bf c}, an ellipsoidal smectic vesicle~\cite{jia}.   Right: Simulation results for a case of $K_3/K_1 \approx 2.0$; {\bf d}, $\kappa = 0.04$; {\bf e}, $\kappa = 0.1$; {\bf f}, $\kappa = 0.5$.   The contour plots show the distribution of the local Frank free energy. }
\label{comparison_exp}
\end{figure}

More prominent shape changes for vesicles with the anisotropic Frank elastic constants are clearly confirmed from a quantitative measurement of the asphericity (i.e., degree of deviation from the reference unit sphere geometry), which is defined as follows:
\be \frac{\langle \Delta R^2\rangle}{R^2_{\rm ref}}=\frac{1}{N} \sum_{\alpha}^N
\frac{(R_{\alpha}-R_{\rm ref})^2}{R_{\rm ref}^2} \ , \ee where $R_{\alpha}$ is
the radial distance of vertex $\alpha$, $R_{\rm ref}$ is the radius of
the reference unit sphere, and N is the total number of vertices.
The asphericities are averaged over 10 simulation runs for each $\kappa$ and plotted in Fig.~\ref{simulation} as a function of $1/{\kappa}$.  The plot exhibits relatively large deviations from its average values, especially at low bending
rigidity. This is mainly attributable to the differences in the asphericity between the three possible ground
state morphologies.
Although faceted tetrahedral vesicles are expected to be the ground state for large system sizes, we have observed in our simulation three different ground state morphologies, presumably due to its finite system size: i) an ellipsoidal vesicle with two closely bounded  disclination pairs; ii) a flattened (square cushion-shape) vesicle with four +1/2 disclinations located approximately in one plane; and iii) a tetrahedral vesicle with four well separated +1/2 defects (see supplementary information).   These three morphologies seem degenerate as the differences in their total free energies are within $0.5\%$.  These vesicle shapes can be viewed as the precursors of the extreme morphologies at $\kappa \rightarrow 0$, such as long fibrous cylinders, double layer sheets, and sharply faceted tetrahedrons, respectively. Finally, we briefly compare our simulation results with our recent experimental
observations in Fig.~\ref{comparison_exp}.

In conclusion, we have studied the fascinating morphology of nematic/smectic vesicles, such as the faceted tetrahedron, nanofibers, and ellipsoids.  Our theoretical and numerical studies provide the fundamental understanding of formation of these novel structured vesicles and elucidate the shape-selective mechanisms.  It could also pave the way for formulating guiding principles in designing nanocarriers with specific shapes, particularly utilizing the two-dimensional nematic order and the topological defects, which are ubiquitous in closed vesicles.

\vspace{0.5cm}

\noindent{\bf Methods}

\noindent To implement a deformable lattice model with spherical topology, we first introduce a reference sphere and tessellate it with a triangular mesh along with 12 requisite 5-disclinations.  Afterward a dual lattice of the triangular mesh is constructed, each dual site being the center of mass of each plaquette formed by the original triangular lattice. The details of the lattice geometry are illustrated in the supplementary information. Let $\hat{m}_{\alpha}$ to be the unit vector normal to the plaquette $\alpha$.  A director  $\hat{n}_{\alpha}$ and a projection operator $\hat{N}_{\alpha} = \hat{n}_{\alpha}\hat{n}_{\alpha}$ are defined on each dual site with a constraint that it must be perpendicular to the plaquette normal: $\hat{N}_{\alpha} \cdot \hat{m}_{\alpha} = 0$.

Let $d_{\alpha\beta}$ be the bond length connecting two neighboring dual sites $\alpha$ and $\beta$, and $S_{\alpha\beta}$  be the area spanned by the bond $\alpha\beta$.  The discretized Frank free energy is then given by
 \begin{eqnarray} F_{\rm Frank}=
 K \sum_{<\alpha\beta>} S_{\alpha\beta} d_{\alpha\beta}^{-2}
\Tr[(\hat{N}_{\beta}-\hat{N}_{\alpha})^2]
\end{eqnarray}
The bond lengths $d_{\alpha\beta}$ and the areas $S_{\alpha\beta}$ are introduced to ensure that the lattice model is invariant under retriangularization, a necessary requirement for modeling a fluid membrane.  In the continuous limit, the retriangularization invariance reduces to the usual reparameterization invariance of a fluid membrane.

The discretized bending energy is given by
\be
F_{\rm bending}= {\kappa} \sum_{\alpha} S_{\alpha}
\Tr  \mathbf{K}_{\alpha}^2 ,
\ee
where $\mathbf{K}_{\alpha}$ is the extrinsic curvature tensor at site $\alpha$, whilst $S_{\alpha}$ is the area of the plaquette $\alpha$.  The curvature tensor $\mathbf{K}_{\alpha}$ of each plaquette $\alpha$ can be calculated from the following three equations:
\begin{eqnarray}
e^{||}_{\alpha\beta} &=& \frac{1}{2}
\vec{e}^{\bot}_{\alpha\beta}\cdot \mathbf{K}_{\alpha} \cdot
\vec{e}^{\bot}_{\alpha\beta}\\\nonumber
e^{||}_{\alpha\gamma} &=& \frac{1}{2}
\vec{e}^{\bot}_{\alpha\gamma}\cdot \mathbf{K}_{\alpha} \cdot
\vec{e}^{\bot}_{\alpha\gamma}\\\nonumber
e^{||}_{\alpha\delta} &=& \frac{1}{2}
\vec{e}^{\bot}_{\alpha\delta}\cdot \mathbf{K}_{\alpha} \cdot
\vec{e}^{\bot}_{\alpha\delta},
\end{eqnarray}
 where $\vec{e}_{\alpha\beta}$ is the vector pointing from vertex $\alpha$ to vertex $\beta$, and $e^{||}_{\alpha\beta}$ and  $e^{\bot}_{\alpha\beta}$ are its components parallel and perpendicular to the plaquette normal $\hat{m}_{\alpha}$.

In the MC simulations, the deformable surface consists of 300 vertices, corresponding to 596 directors in all. The initial shape of the  surface is a unit sphere and the initial director orientations are random.
Each MC sweep consists of  trial attempts to rotate each director and to move each vertex.  The acceptance or rejection of a MC trial is determined by the standard Metropolis algorithm.   All vertices are allowed to move along the radial direction with the angular positions of the vertices fixed.  In order to preserve the total area upon surface deformation, any vertex moves making a total area change larger than $1\%$ are rejected.  Finally, once the surface is deformed by vertex moves, the orientations of directors are corrected by projecting them on the newly deformed plaquette before the new free energy is calculated.

\noindent {\bf Acknowledgements}
XX thanks Shanghai Jiao Tong University for financial support.  The work of HS was supported by the NSF through grant DMR 09-55760.  The work of MJB and ZY was supported by the NSF through grant DMR-0808812 and that of ZY by funds from Syracuse University. The work of LJ and MHL was supported by the French ANR grant ANR-08-BLANC-0209-01.   

\noindent{\bf Author contributions}
XX and MJB designed the project.  XX, MJB and ZY developed the theory.  HS developed and executed the numerical model.  LJ and MHL designed and performed experiments.  All author contributed to the writing of the paper.

\end{document}